\documentclass[a4paper,fleqn,UKenglish]{article}

\usepackage{amsthm,amssymb}
\usepackage[margin=2.4cm]{geometry}
\usepackage[T1]{fontenc} 
\usepackage[utf8]{inputenc} 
\usepackage{lmodern} 
\usepackage{microtype}
\usepackage{cite}
\usepackage{ellipsis} 
\usepackage{xspace}
\usepackage{enumitem}
\usepackage{mathtools}
\usepackage{dsfont}
\usepackage{algorithm}
\usepackage[noend]{algpseudocode}
\usepackage{tikz}
\usepackage{xcolor}
\usetikzlibrary{positioning}
\usetikzlibrary{shapes.multipart}
\usepackage{graphicx}
\usepackage{authblk}

\usepackage{lettrine}
\usepackage{booktabs}
\usepackage{hyperref}
\usepackage[capitalize]{cleveref}

\newtheorem*{fact}{Fact}

\crefname{enumi}{}{}

\newcommand{\problem}{\textsc{Online Unbounded Knapsack Problem with Removal}\xspace}

\newcommand{\euler}{\mathrm{e}}
\newcommand{\ie}{that is}
\newcommand{\eps}{\varepsilon}
\renewcommand{\P}{\normalfont\text{Prob}}
\newcommand{\E}{\mathds{E}}
\newcommand{\alg}{{\normalfont\textsc{Alg}}\xspace}
\newcommand{\opt}{{\normalfont\textsc{Opt}}\xspace}

\newcommand{\genalg}{{\normalfont\textsc{Focus}}\xspace}
\newcommand{\genbold}{\textnormal{\textbf{\textsc{Focus}}}\xspace}
\newcommand{\propalg}{{\normalfont\textsc{Simple}}\xspace}
\newcommand{\propbold}{\textnormal{\textbf{\textsc{Simple}}}}
\newcommand{\proprandalg}{{\normalfont\textsc{RandChoice}}\xspace}
\newcommand{\proprandbold}{\textnormal{\textbf{\textsc{RandChoice}}}\xspace}
\newcommand{\multset}[1]{\text{fill}(#1)}

\newcommand{\sylv}[1]{a_{#1}}

\newcommand{\upsum}[1]{S_{#1}}

\newcommand{\sylvsum}{S_{\infty}}
\newcommand{\partsylv}[1]{T_{#1}}
\newcommand{\longcomment}[1]{}

\newcommand{\N}{\mathds{N}}
\newcommand{\R}{\mathds{R}}
\newcommand{\Rplus}{\R_{> 0}}
\renewcommand{\phi}{\varphi}

\crefname{equation}{Equation}{Equations}

\newtheorem{lemma}{Lemma}

\newtheorem{theorem}{Theorem}

\title{Stealing From the Dragon's Hoard\thanks{A previous version of this paper appeared at STACS 2026~\cite{GS2026}.}\\

\smallskip

\large Online Unbounded Knapsack With Removal}

\author[1]{Matthias Gehnen}
\author[2]{K\"ubra G\"uven}
\author[3]{Moritz Stocker\thanks{Corresponding author}}

\affil[1]{Department of Mathematics and Computer Science, Syddansk Universitet, Denmark \authorcr {\small\texttt{gehnen@imada.sdu.dk}}}
\affil[2]{Department of Computer Science, RWTH Aachen University, Germany \authorcr {\small\texttt{kuebra.gueven@rwth-aachen.de}}}
\affil[3]{Department of Computer Science, ETH Zurich, Switzerland \authorcr {\small\texttt{moritz.stocker@inf.ethz.ch}}}
\date{}

\begin{document}
\maketitle

\bigskip

\begin{abstract}
We introduce the \textsc{Online Unbounded Knapsack Problem with Removal}, a variation of the well-known \textsc{Online Knapsack Problem}. Items, each with a weight and value, arrive online and an algorithm must decide on whether or not to pack them into a knapsack with a fixed weight limit. An item may be packed an arbitrary number of times and items may be removed from the knapsack at any time without cost. The goal is to maximize the total value of items packed, while respecting the weight limit. We show that this is one of the very few natural online knapsack variants that allow for competitive deterministic algorithms in the general setting, by providing an algorithm with competitivity $\sylvsum\approx 1.691$. We complement this with a matching lower bound.

We also analyze the proportional setting, where the weight and value of any single item agree, and show that deterministic algorithms can be exactly $3/2$-competitive. Lastly, we give lower and upper bounds of $6/5$ and $4/3$ on the competitivity of randomized algorithms in this setting.
\end{abstract}

\section{Introduction}\label{sec:introduction}

\begin{quote}
\lettrine[lines=3,slope=-0.5em,lhang=0.1]{\fontfamily{yfrak}\textbf{I}}{magine,}  if you will, a hobbit creeping through a dragon's lair. Having found the dragon, 
it now swiftly makes its escape. Surrounding it on its way out are piles of coins, heaps of rubies, mountains of diamonds, each containing far more than the hobbit could possibly carry. Grinning, the hobbit opens its trusty knapsack and fills it to the brim with silver coins. Around a corner, it almost stumbles over a number of large gold bars. It empties its knapsack of silver and packs two of these gold bars instead, which completely fill its knapsack.
Around the next bend, gleaming in the light of the hobbit's candle, lies an enormous jewel.
Greedily, the hobbit removes the two gold bars and packs the jewel instead. It silently curses, wishing it had kept the silver coins, some of which would have nicely filled up the remaining space in the knapsack. But it dares not go back, deeper into the lair, towards the dormant dragon. And thus, the hobbit shoulders its knapsack and presses on\dots
\end{quote}
In the \textsc{Online Knapsack Problem}, an algorithm receives an unknown number of items one by one, each with a weight and a value. It must decide whether or not to pack these items into a knapsack, without any knowledge about future items that may be offered. Its goal is to maximize the combined value of the items it packs without exceeding the knapsack's weight capacity, often normalized to $1$. 

Ever since the problem was introduced by Marchetti-Spaccamela and Vercellis~\cite{MV1995}, it has offered fertile ground for analysis. Though the original problem is not competitive, even for the \textsc{Proportional Knapsack} variant where the weight and value of any single item are equal, there are countless variants of the problem and aspects under which to consider them that prove interesting---and for which well-performing algorithms exist. A recent survey detailing many of these results
has been compiled by Böckenhauer et al.~\cite{BHKRS2026}.

One of the simplest competitive variants is the \textsc{Online Knapsack Problem with Removal}. In this model, items may be removed from the knapsack at any time without cost. It was first studied for the proportional case by Iwama and Taketomi~\cite{IT2002}, who showed that the best competitive ratio that can be achieved by a deterministic algorithm is exactly the golden ratio $\phi\approx 1.618$. Han et al.~\cite{HKM2015} considered randomized algorithms and provided an algorithm with competitivity $10/7$. They also gave a lower bound of $5/4$ on the competitive ratio of any randomized algorithm, which has recently been improved by Hächler~\cite{Haechler2025} to approximately $1.2695$.
The general case, where items have arbitrary weights and values, remains non-competitive for deterministic algorithms even with removal, as shown by Iwama and Zhang~\cite{IZ2010}. However, Han et al.~\cite{HKM2015} showed that a competitivity of $2$ can in fact be achieved by randomized algorithms, which they complemented by a lower bound of $1+1/\euler$.

More recently, Böckenhauer et al.~\cite{BGHKKLMRS2024} considered the \textsc{Online Unbounded Knapsack Problem}. This can be seen as the ``industrial'' setting of the knapsack problem, where each item is available in unlimited quantities. In this model, rather than only being able to decide on whether to pack an item or not, an algorithm may choose to pack it any number of times. Correspondingly, the classical model, where each item can only be packed once or not at all, is sometimes referred to as the $0$-$1$ \textsc{Knapsack Problem}. More correctly, the \textsc{Unbounded Knapsack Problem} should not be viewed as a model that provides additional resources to an algorithm, but as one that limits the possible \emph{instances} considered. An instance of the \textsc{Unbounded Knapsack Problem} is equivalent to an instance of the $0$-$1$ \textsc{Knapsack Problem} where any item of weight $w$ is guaranteed to consecutively appear at least $\lfloor 1/w\rfloor$ times. Böckenhauer et al.\ showed that deterministic algorithms can be exactly $2$-competitive in the proportional case, and gave lower and upper bounds of $1+\ln(2)\approx 1.6931$ and $1.7353$, respectively, for randomized algorithms. They also showed that in the general setting, no finite competitive ratio is possible even for randomized algorithms.

\paragraph*{Our Contributions}
In this paper, we combine the \textsc{Online Unbounded Knapsack Problem} and the \textsc{Online Knapsack Problem with Removal} to the \textsc{Online Unbounded Knapsack Problem with Removal}. 
This can be interpreted as an application of removal, being one of the most natural variants of the knapsack problem, to the industrial setting of the Unbounded Knapsack, resulting in a situation where an algorithm 
is not penalized for initially being forced to pack worthless items.

More whimsically, it can be described by a hobbit stealing from the riches of a dragon's hoard. The online nature of the problem is captured by the fact that the hobbit's first priority is to move away from the dragon, while it tries to maximize the value of items it takes along. It is natural to assume that the hobbit would be able to remove items from its knapsack. Given that a dragon's hoard is commonly associated with untold riches in vast quantities, the unbounded setting is also a reasonable choice.

In this model, we show that the best possible competitive ratio for deterministic algorithms in the proportional variant is exactly $3/2$ (\cref{lemma:proplower,lemma:propupper}). For randomized algorithms, we offer a simple lower bound of $6/5$ on the competitive ratio (\cref{lemma:proprandlower}) and a slightly more complex randomized algorithm with competitivity $4/3$ (\cref{thm:proprandupper}).
These results are summarized in \cref{table:results}.
This model is obviously no harder than the \textsc{Online Unbounded Knapsack Problem} (since it lends additional capabilities to the algorithm) and the \textsc{Online Knapsack Problem with Removal} (since it essentially limits the set of possible instances), so it is not surprising that all our bounds are lower than the best-known bounds in those models.

Our main results concern the \textsc{General Knapsack Problem}, however. Interestingly, the \problem is one of the few natural knapsack variants where deterministic algorithms can be competitive without further assumptions.\footnote{The few other problems that allow for this include \textsc{Online Knapsack with Removal} with resource augmentation~\cite{IZ2010} and resource buffer~\cite{HKMY2019}, as well as \textsc{Online Multiple Knapsack with Removal}~\cite{CJS2016}, all of which tinker with the size of the knapsack and are arguably less simple than the variant we consider.}
Neither the \textsc{Online Knapsack Problem with Removal} alone~\cite{IZ2010} nor the \textsc{Online Unbounded Knapsack Problem} (without removal)~\cite{BGHKKLMRS2024}  allow for this behavior. 
However, we show that in the combined model of the \problem, a simple algorithm can achieve a competitivity of $\sylvsum\approx 1.6910$ (\cref{thm:genupper}). We complement this with a matching lower bound (\cref{thm:genlower}).

Interestingly, this is not the first time that $\sylvsum$ has appeared as the competitive ratio of an online algorithm for a packing problem: it was previously shown to be the (asymptotic) competitive ratio of the \textsc{Harmonic} algorithm for \textsc{Bin-Packing}~\cite{LL1985}. The lower bound of \cref{thm:genlower} shows, however, that the speculation regarding a connection between the problem considered in this paper and \textsc{Bin-Packing} voiced in the previous version~\cite{GS2026} is unfounded.

\begin{table}[t]
\caption{Results contained in this paper (left column), as compared to previous results for \textsc{Online Unbounded Knapsack} and \textsc{Online Knapsack with Removal}. These results cover deterministic and randomized algorithms for the proportional case, as well as deterministic algorithms for the general case. The value $\sylvsum$ satisfies $\sylvsum\approx 1.6910$. Randomized algorithms for the general problem are not studied separately in this paper; the lower bound follows from the proportional case, while the upper bound follows from the deterministic case.}
\label{table:results}

\medskip 

\begin{tabular}{lcccccc}
\toprule
&\multicolumn{2}{c}{\textsc{Unbounded with Removal}}&\multicolumn{2}{c}{\textsc{Unbounded} \cite{BGHKKLMRS2024}}&\multicolumn{2}{c}{\textsc{With Removal}}\\
\midrule
&lower bound&upper bound& l.b. & u.b.& l.b. & u.b.\\
\midrule
prop.\ det. & $3/2$  {\footnotesize(Lemma \ref{lemma:proplower})} & $3/2$ {\footnotesize(Lemma \ref{lemma:propupper})} & \multicolumn{2}{c}{$2$}&\multicolumn{2}{c}{$1.618$ \cite{IT2002}}\\
prop.\ rand. & $6/5$ {\footnotesize(Lemma \ref{lemma:proprandlower})} & $4/3$ {\footnotesize(Thm.\ \ref{thm:proprandupper})} & $1.6931$ & $1.7353$ & $1.2695$\cite{Haechler2025} & $10/7$\cite{HKM2015}\\
general det. & $\sylvsum$ {\footnotesize(Thm.\ \ref{thm:genlower})} & $\sylvsum$ {\footnotesize(Thm.\ \ref{thm:genupper})} & \multicolumn{2}{c}{$\infty$}&\multicolumn{2}{c}{ $\infty$\cite{IZ2010}}\\
general rand. & $6/5$ {\footnotesize(Lemma \ref{lemma:proprandlower})} & $\sylvsum$ {\footnotesize(Thm.\ \ref{thm:genupper})} & \multicolumn{2}{c}{$\infty$} & $1.3679$~\cite{HKM2015} & $2$~\cite{HKM2015}\\
\bottomrule
\end{tabular}
\end{table}

\paragraph*{Preliminaries}
An instance of the \textsc{Online Unbounded Knapsack Problem with Removal} consists of a sequence of items $I=(x_1,\dots,x_n)$. 
An item is given as $x=(w,v)$, where $w=w(x)\in \left]0,1\right]$ is its \emph{weight} and $v=v(x)\in \Rplus$ is its \emph{value}; we define the item's \emph{density} as $\rho(x)=v(x)/w(x)$.

A solution is given as a multiset $S$ of items from $I$, where each item can appear an arbitrary number of times, satisfying
\[\sum_{x\in S} w(x)\leq 1\,.\]
The goal is to maximize the \emph{gain}
\[v(S)=\sum_{x\in S} v(x)\,.\]
We denote by $\opt(I)$ an optimal solution to $I$, \ie, a solution that maximizes $v(\opt(I))$. 

An online algorithm \alg maintains a solution (or ``knapsack'') $S$. When it receives an item $x$, it may add (or ``pack'') any number of copies of $x$ to $S$, then remove an arbitrary subset of items from $S$, such that $S$ is again a valid solution. We denote by $\alg(I)$ the solution of $\alg$ on the instance $I$.

A deterministic algorithm \alg is said to be (strictly) $c$-competitive for some constant $c\geq 1$ if $c\cdot v(\alg(I))\geq v(\opt(I))$ for any instance $I$. Equivalently, a randomized algorithm \alg is said to be (strictly) $c$-competitive in expectation if $c\cdot \E[v(\alg(I))]\geq v(\opt(I))$ for any instance $I$.
In either case, the \emph{(strict) competitive ratio} or \emph{competitivity} of \alg is defined as
\[\inf_{c\geq 1} \{c\mid\text{\alg is (strictly) $c$-competitive (in expectation)}\}\,.\]

There is also a non-strict version of competitivity: an algorithm is said to be (non-strictly) $c$-competitive if there exists a constant $\alpha\geq 0$ such that $c\cdot v(\alg(I))+\alpha\geq v(\opt(I))$ for any instance $I$.
Throughout this paper, we only consider the strict competitive ratio and reserve the term \emph{competitive ratio} for this version. It should be noted, however, that the lower bound for general item values of \cref{thm:genlower} also holds for the non-strict version, which can easily be seen by scaling the values of all items.  For any upper bound, proving a strict competitive ratio obviously implies a corresponding non-strict ratio.

For an item $x=(w,v)$, the maximal number of copies of $x$ that fit into the knapsack at the same time is exactly $\lfloor 1/w(x)\rfloor$. We define the \emph{cumulative value} 
$v^*(x)=v(x)\cdot \lfloor 1/w(x)\rfloor$
as the combined value of these copies, \ie, the maximal gain that can be achieved using only copies of $x$. Similarly, we define the \emph{cumulative weight} of $x$ as $w(x)\cdot \lfloor 1/w(x)\rfloor$. We denote by $\multset{x}$ the corresponding multiset consisting of $\lfloor 1/w(x)\rfloor$ copies of $x$.

We finally state a simple result that will prove useful throughout the paper:
\begin{lemma}
\label{lemma:multiplicity}
If $0<x\leq 1/n$, for $n\in\mathbb{N}$ and $n\geq 1$, then 
\[x\cdot \lfloor 1/x\rfloor=\frac{\lfloor 1/x\rfloor}{1/x}\geq \frac{n}{n+1}\,.\]
\end{lemma}
\begin{proof}
Let $n'=\lfloor 1/x\rfloor$, so $1/x\leq n'+1$. Since $1/x\geq n$, clearly $n'\geq n$, and therefore
\[\frac{\lfloor 1/x\rfloor}{1/x}=\frac{n'}{1/x} \geq \frac{n'}{n'+1}\geq \frac{n}{n+1}\,.\]
\end{proof}

\section{The Proportional Knapsack Problem}

In this section, we consider the \textsc{Proportional Knapsack Problem}, where the weight and value of any item are equal. For convenience, we identify an item $x$ with its weight and value, \ie, we say that $x=w(x)=v(x)$.

\subsection{Deterministic Algorithms}\label{subsec:propdet}

\textbf{The algorithm \propbold} keeps the heaviest item in the instance, as long as there are no items of weight less than or equal to $1/2$. If there is an item of weight smaller than or equal to $1/2$ in the instance, \propalg removes any previous item in the knapsack, packs as many copies of this small item as possible and ignores all further items.

The algorithm is shown in pseudocode as \cref{alg:detalg}.

\begin{algorithm}[tp]
\caption{The deterministic algorithm \propalg for the proportional knapsack problem}\label{alg:detalg}
\begin{algorithmic}
\State $K\gets \emptyset$
\For{$x\in I$}
	\If{$x\leq 1/2$}
		\State $K\gets \multset{x}$
		\State \Return $K$
	\ElsIf{$K=\emptyset$}
		\State{$K\gets \{x\}$}
	\ElsIf{$K=\{x'\}$ and $x>x'$}
		\State{$K\gets \{x\}$}
	\EndIf
\EndFor
\State \Return $K$
\end{algorithmic}
\end{algorithm}

\begin{lemma}\label{lemma:propupper}
The competitive ratio of \propalg is at most $3/2$. 
\end{lemma}

\begin{proof}
Consider any instance $I$. If there are no items of weight less than or equal to $1/2$ in $I$, no two items in the instance fit in the knapsack together. Since \propalg keeps the heaviest item, it is optimal.

Otherwise, let $x$ be the first item in $I$ of weight $x\leq 1/2$. By definition, \propalg achieves a gain of $x\cdot \lfloor 1/x\rfloor$, which is at least $2/3$ by \Cref{lemma:multiplicity}. Since the optimal solution has value at most $1$, \propalg therefore has a competitive ratio of at most $3/2$. 
\end{proof}

While, as its name indicates, \propalg uses a straightforward strategy, we can show that no deterministic online algorithm can improve on its competitive ratio.

\begin{lemma}\label{lemma:proplower}
No deterministic algorithm for the \textsc{Proportional Online Unbounded Knapsack Problem with Removal} can have a competitive ratio of less than $3/2$.
\end{lemma}

\begin{proof}
Let \alg be any online algorithm and let $\eps>0$ be sufficiently small.
Consider the two instances
\begin{align*}
I_1&= (1/3+2\eps,2/3-\eps,1/3+\eps)\,,\\
I_2&=(1/3+2\eps,2/3-\eps,2/3-2\eps)\,.
\end{align*}

Since $I_1$ and $I_2$ agree on the first two items, \alg must act identically on that prefix.
If it packs the second item $2/3-\eps$, it cannot keep any copies of $1/3+2\eps$. In this case, it has a gain of at most $2/3-\eps$ on $I_2$, while by packing one copy of $1/3+2\eps$ and the last item, a gain of $1$ would have been possible.

If, on the other hand, it does not pack the item $2/3-\eps$, it cannot pack more than two copies of $1/3+2\eps$ and $1/3+\eps$ in total. Therefore, its gain on $I_1$ is at most $2/3+4\eps$, while again, a gain of $1$ is possible by packing the second item and one copy of the last item.
In either case, it achieves a competitive ratio of at least $1/(2/3+4\eps)$. The statement of the lemma follows as $\eps\to 0$.
\end{proof}

\subsection{Randomized Algorithms}

We now consider randomized algorithms for the \textsc{Proportional Online Unbounded Knapsack Problem with Removal}. 
We start with a simple lower bound using Yao's principle~\cite{Yao1977}:

\begin{fact}[Yao's Principle]\label{fact:yao}
  Consider any online maximization problem, where $v(S)$ denotes the value of a solution $S$. Let
  $\mathcal{I}$ be a set of instances, and let $\P_{\mathcal{I}}$ denote any
  probability distribution on $\mathcal{I}$, with $\E_{\mathcal{I}}$ being the corresponding expectation. If, for any deterministic algorithm \alg, it holds that 
  \[c\cdot \E_{\mathcal{I}}[v(\alg)]\leq \E_{\mathcal{I}}[v(\opt)]\,,\] for some constant $c\geq 1$, then no randomized algorithm for the problem can have a competitive ratio of less than $c$.
\end{fact}

A more detailed explanation of Yao's principle can be found, for example, in the textbook by Komm~\cite{Komm2016}.

\begin{lemma}\label{lemma:proprandlower}
No randomized algorithm for the \textsc{Proportional Online Unbounded Knapsack Problem with Removal} can have a competitive ratio of less than $6/5$. 
\end{lemma}

\begin{proof}
The proof uses the same instances as in \Cref{lemma:proplower}, that is,
\begin{align*}
I_1&= (1/3+2\eps,2/3-\eps,1/3+\eps)\,,\\
I_2&=(1/3+2\eps,2/3-\eps,2/3-2\eps)\,,
\end{align*}
for some sufficiently small $\eps>0$. Let \alg be any deterministic algorithm that receives $I_1$ or $I_2$ each with probability $1/2$. Then, as in the proof of \Cref{lemma:proplower}, if \alg packs the item $2/3-\eps$, its gain is at most $2/3-\eps$ on $I_2$ and at most $1$ on $I_1$. 
If it does not pack the item $2/3-\eps$, its gain is at most $2/3+4\eps$ on $I_1$ and at most $1$ on $I_2$. In either case, its expected gain is at most $1/2\cdot (2/3+4\eps+1)=5/6+2\eps$, while the optimal solution is always $1$. Therefore, by Yao's principle, no randomized algorithm can have a competitive ratio of less than $(5/6+2\eps)^{-1}$. The statement of the lemma follows as $\eps\to 0$.
\end{proof}

We now present a randomized algorithm for the \textsc{Proportional Online Unbounded Knapsack Problem with Removal} that uses a single random bit and has a competitive ratio of at most $4/3$.

\textbf{The algorithm} \proprandbold divides items into four different weight categories. \emph{Good} items $G=\left[0,1/3\right]\cup \left[3/8,1/2\right]\cup \left[3/4,1\right]$, \emph{small} items $S=\left]1/3,3/8\right[$, \emph{medium} items $M=\left]1/2,5/8\right[$, and \emph{large} items $L=\left[5/8,3/4\right[$. It chooses uniformly at random between two deterministic strategies $A_1$ and $A_2$. In both strategies, if it encounters an item in $G$, it removes any items in the knapsack, packs the new item as often as possible and ignores all further items. Otherwise the corresponding strategies are as follows:

\begin{itemize}
\item \mbox{\boldmath$A_1\colon$\unboldmath} \proprandalg initially keeps copies of a single item, in the following order of priority:
\begin{enumerate}[left=1.5em]
\item Two copies of the smallest item in $S$ encountered so far,
\item The largest item in $M$ encountered so far (if none in $S$ are present), 
\item The smallest item in $L$ encountered so far (if none in $S\cup M$ are present).
\end{enumerate}
To do so, it removes any previous items in the knapsack if necessary.

If at any point it encounters an item from $S$ that fits with an item in $M\cup L$ that is already in the knapsack or vice versa, it packs one copy of each. From then on, it replaces the item in $S$ with any smaller item from $S$ if possible, and the item from $M\cup L$ with a larger item from $M\cup L$ if it can do so without removing the item from $S$. 
\item \mbox{\boldmath$A_2\colon$\unboldmath} \proprandalg initially keeps copies of a single item, in the following order of priority:
\begin{enumerate}[left=1.5em]
\item The smallest item in $L$ encountered so far,
\item Two copies of the smallest item in $S$ encountered so far (if none in $L$ are present),
\item The largest item in $M$ encountered so far (if none in $S\cup L$ are present).
\end{enumerate}
To do so, it removes any previous items in the knapsack if necessary.

If at any point it encounters an item from $S$ that fits with an item in $L$ (but not $M$) that is already in the knapsack or vice versa, it packs one copy of both and ignores all further items in $S\cup M \cup L$. 
\end{itemize}

The algorithm is shown in pseudocode as \Cref{alg:proprandalg}.

\begin{theorem}\label{thm:proprandupper}
\proprandalg achieves a competitive ratio of at most $4/3$. 
\end{theorem}

\begin{proof}
Let $I$ be any non-empty instance and let $\opt\coloneqq\opt(I)$ be a fixed optimal solution of value $v(\opt)$.
We first consider what happens when there is an item from $G$ in the instance. Let $g\in G$ be the first such item. In that case, \proprandalg achieves a gain of $g\cdot \lfloor 1/g\rfloor$. If $g\in [0,1/3]$, this is at least $3/4$ by \Cref{lemma:multiplicity}. If $g\in [3/8,1/2]$, we know that $g\cdot \lfloor 1/g\rfloor=2g\geq 3/4$. If $g\in [3/4,1]$, \proprandalg also achieves a gain of at least $3/4$. In any case, its competitive ratio is at most $4/3$. We can therefore assume that $I$ only contains items in $S\cup M\cup L$. Additionally, if $I$ contains only items from $M$, then \proprandalg is clearly optimal since both strategies keep the largest such item in this case. We therefore also assume that $I$ contains an item from $S\cup L$. 

If $v(\opt)\leq 3/4$, then \proprandalg is easily seen to be $4/3$-competitive: Strategy $A_1$ always achieves a gain of at least $1/2$ and, since the instance contains an item from $S\cup L$, Strategy $A_2$ achieves a gain of at least $5/8$ (any item from $S$ has a cumulative value of at least $2/3> 5/8$). The expected gain of \proprandalg is therefore at least $1/2\cdot (1/2+5/8)=9/16$, and its competitive ratio in expectation is at most $(3/4)/(9/16)=4/3$. We can therefore assume that $v(\opt)>3/4$, which means that $\opt$ contains one copy of an item $s\in S$ and one copy of an item $b\in M\cup L$. We will now distinguish three different cases:
\begin{itemize}[itemsep=0.4em]
\item \mbox{\bfseries\boldmath Case 1 ($b\in M$):\unboldmath} Since there is an item from $S\cup L$ in the instance, as above, Strategy $A_2$ will achieve a gain of at least $5/8\geq 5/8\cdot v(\opt)$. 
Since any item from $S$ fits with any item from $M$, and since it prioritizes items from $S\cup M$ over items from $L$, Strategy $A_1$ will pack both an item from $S$ and an item from $M\cup L$. 
More specifically, since it first prioritizes larger items from $M$ and then replaces the item from $M\cup L$ with larger items if possible, it packs an item from $S$ and an item of size at least $b$, leading to a gain of at least $1/3+b$. On the other hand, $v(\opt)\leq 3/8+b$. Therefore, Strategy $A_1$ achieves a gain of at least $(1/3+b)/(3/8+b)\cdot v(\opt)$. Since $(1/3+b)/(3/8+b)$ increases with $b$, the gain of Strategy $A_1$ is at least $(1/3+1/2)/(3/8+1/2)\cdot v(\opt)=20/21\cdot v(\opt)$.
Combining Strategies $A_1$ and $A_2$, the algorithm \proprandalg has an expected gain of $1/2\cdot (5/8+20/21)\cdot v(\opt)=265/336\cdot v(\opt)>3/4\cdot v(\opt)$.
\item \mbox{\bfseries\boldmath Case 2 ($b\in L$ and $s$ arrives before $b$):\unboldmath} As in the previous case, Strategy $A_2$ achieves a gain of at least $5/8$. By definition of Strategy $A_1$, it will have packed an item from $S$ of size at most $s$ when $b$ arrives, possibly together with an item from $M\cup L$. This means that it could pack $b$ together with that item, and since it replaces the item from $M\cup L$ with a larger item if possible, it has a gain of at least $1/3+b\geq 1/3+5/8= 23/24$. This means that \proprandalg has an expected gain of at least $1/2\cdot (5/8+23/24)=19/24>3/4$.
\item \mbox{\bfseries\boldmath Case 3 ($b\in L$ and $b$ arrives before $s$):\unboldmath} Since the instance contains an item from $S$, Strategy $A_1$ achieves a gain of at least $2/3$. Strategy $A_2$ on the other hand is always able to pack both an item from $S$ and an item from $L$: if it still has only one item packed when $s$ arrives, that item must be an item from $L$ of size at most $b$. Therefore, Strategy $A_2$ achieves a gain of at least $1/3+5/8=23/24$ and \proprandalg achieves an expected gain of at least $1/2\cdot (2/3+23/24)=13/16>3/4$.
\end{itemize}

In any case, \proprandalg has an expected gain of at least $3/4\cdot v(\opt)$ and thus a competitive ratio in expectation of at most $4/3$.
\end{proof}

It may seem that the competitive ratio of \proprandalg can easily be improved by not choosing the two strategies uniformly. The competitive ratio of either strategy is, however, bounded from below by $4/3$ by the assumptions made before the case distinction and especially by the way they deal with items from $G$. For example, if the first item \proprandalg receives has weight $3/4$, it will pack it and ignore all future items, while the optimal solution might well be $1$.

Though it may be possible to improve the algorithm, this would thus require a more fine-grained treatment of the items in $G$.

\begin{algorithm}[p]
\caption{The randomized algorithm \proprandalg for the proportional knapsack problem}\label{alg:proprandalg}
\begin{algorithmic}
\State $\parbox{1em}{$G$}\gets[0,1/3]\cup [3/8,1/2]\cup [3/4,1]$
\State $\parbox{1em}{$S$}\gets\;]1/3,3/8[$
\State $\parbox{1em}{$M$}\gets\;]1/2,5/8[$
\State $\parbox{1em}{$L$}\gets[5/8,3/4[$
\State $\parbox{1em}{$K$}\gets \emptyset$
\State $\parbox{1em}{$b$}\gets \text{rand}(\{0,1\})$
\Comment{Uniformly random choice between two strategies}
\If{$b=1$}
\Comment{Deterministic strategy $A_1$}
	\For{$x\in I$}
		\If{$x\in G$}
			\State $K\gets \multset{x}$
			\State \Return $K$
		\ElsIf{$K=\emptyset$}
			\State{$K\gets \multset{x}$}
		\ElsIf{$K=\multset{x'}$ for $x'\in S\cup M \cup L$}
			\If{$\min(x,x')\in S$ and $\max(x,x')\in M\cup L$ and $x+x'\leq 1$}
				\State $K\gets \{x,x'\}$
			\ElsIf{$x\in S$ and $x<x'$}
				\State $K\gets \{x,x\}$
			\ElsIf{$x\in M$ and $x'\in M$ and $x>x'$}
				\State $K\gets \{x\}$ 
			\ElsIf{$x'\in L$ and $x<x'$}
				\State $K\gets \{x\}$
			\EndIf
		\Else
			\State $K=\{s,b\}$ with $s\in S$ and $b\in M\cup L$
			\If{$x<s$} 
				\State $K\gets \{x,b\}$
			\ElsIf{$x>b$ and $x+s\leq 1$} 
				\State $K\gets \{s,x\}$
			\EndIf
		\EndIf
	\EndFor
\Else
\Comment{Deterministic strategy $A_2$}
	\For{$x\in I$}
		\If{$x\in G$}
			\State $K\gets \multset{x}$
			\State \Return $K$
		\ElsIf{$K=\emptyset$}
			\State{$K\gets \multset{x}$}
		\ElsIf{$K=\multset{x'}$ for $x'\in S\cup M \cup L$}
			\If{$\min(x,x')\in S$ and $\max(x,x')\in L$ and $x+x'\leq 1$}
				\State $K\gets \{x,x'\}$
			\ElsIf{$x\in L$ and $x'\in S\cup M$}
				\State $K\gets \{x\}$
			\ElsIf{$x\in L$ and $x<x'$}
				\State $K\gets \{x\}$
			\ElsIf{$x\in S$ and $x'\in S\cup M$ and $x<x'$}
				\State $K\gets \{x,x\}$ 
			\ElsIf{$x\in M$ and $x'\in M$ and $x>x'$}
				\State $K\gets \{x\}$
			\EndIf
		\EndIf
	\EndFor
\EndIf
\State \Return $K$
\end{algorithmic}
\end{algorithm}

\section{The General Knapsack Problem}

We now consider the general case of the problem, where the value of an item may be unrelated to its weight. We first show that there are in fact competitive deterministic algorithms for the \textsc{General Online Unbounded Knapsack Problem with Removal}, which may come as a surprise, given that most variants of the \textsc{General Online Knapsack Problem} remain non-competitive~\cite{IZ2010}, many of them even for randomized algorithms~\cite{BKKR2014, BGHKKLMRS2024}. 
To this end, consider an algorithm \genalg.

\textbf{The algorithm} \genbold 
only keeps copies of a single item of maximal cumulative value. If an item of higher cumulative value arrives, it removes all items in the knapsack and packs as many copies of this new item as possible.

The algorithm is shown in pseudocode as \cref{alg:genalg}.
Note that if \genalg is used in an instance of the \textsc{Proportional Online Knapsack Problem}, its gain is at least that of the algorithm \propalg from \cref{subsec:propdet}, which also only keeps copies of a single item. Therefore, \genalg also achieves the best possible competitive ratio of $3/2$ in the proportional case.

\begin{algorithm}[t]
\caption{The deterministic algorithm \genalg for the general knapsack problem}\label{alg:genalg}
\begin{algorithmic}
\State $K\gets \emptyset$
\For{$x\in I$}
	\If{$K=\emptyset$}
		\State $K\gets \multset{x}$
	\ElsIf{$K=\multset{x'}$ and $v^*(x)>v^*(x')$}
		\State{$K\gets \multset{x}$}
	\EndIf
\EndFor
\State \Return $K$
\end{algorithmic}
\end{algorithm}

To analyze the competitivity of \genalg in the general case, we define a sequence of numbers $\sylv{n}$ for $n\geq 1$ recursively by 
\begin{equation}\label{eq:defsylvester}
\sylv{1}=2, \quad \sylv{n}=1+\prod_{i=1}^{n-1} \sylv{i} \quad \text{for }n\geq 2\,.
\end{equation}

The following properties of this sequence\footnote{The sequence, whose first elements are $2,3,7,43,1807,\dots$, is known as \emph{Sylvester's sequence} (sequence A000058 in the OEIS~\cite{oeis}) after James Joseph Sylvester, who studied a family of unit fractions related to it~\cite{Sylvester1880}.} can be easily verified by induction:
\begin{lemma}
\label{lemma:sylvester}
For any $n\geq 1$,
\begin{enumerate}[label=(\roman*)]
\item\label{lemmitem:sylvproduct} $\displaystyle\sylv{n+1}=\sylv{n}\cdot (\sylv{n}-1)+1$\,,
\item\label{lemmitem:sylvremainder} $\displaystyle\frac{1}{\sylv{n}-1}=1-\sum_{j=1}^{n-1}\frac{1}{\sylv{j}}$\,.
\qed
\end{enumerate}
\end{lemma}

Now let $N\geq 1$. We further define 
\begin{align*}
\upsum{N}&=\sum_{n=1}^N (\sylv{n}-1)^{-1},\quad\text{with }\upsum{0}=0\,,\\
\partsylv{N}&=\frac{\sylv{N}}{(\sylv{N}-1)^2}+\upsum{N-1}\,.
\end{align*}

Lastly, we define
$\sylvsum=\sum_{n=1}^\infty  (\sylv{n}-1)^{-1}$; clearly, $\upsum{N}\to \sylvsum$ as $N\to\infty$. Since the numbers $a_n$ grow indefinitely, we also know that $\sylv{N}/(\sylv{N}-1)^2\to 0$, and thus also $\partsylv{N}\to \sylvsum$ as $N\to \infty$. 
 Furthermore, the numbers $\partsylv{N}$ satisfy the following two properties:  
\begin{lemma}
\label{lemma:properties}
For any $N\geq 1$ and $1\leq k\leq N$,
\begin{enumerate}[label=(\roman*)]
	\item\label{lemma:TNone} \(\displaystyle\frac{(\sylv{N}-1)^2}{\sylv{N}}\cdot \bigl(1-\upsum{N-1}/\partsylv{N}\bigr)=\partsylv{N}^{-1}\,.\)
	\item\label{lemma:TNtwo} \(\displaystyle\frac{\sylv{k}\cdot (\sylv{k}-1)}{\sylv{k}+1}\cdot \bigl(1- \upsum{k-1}/\partsylv{N}\bigr)>\partsylv{N}^{-1}\,.\)
\end{enumerate}
\end{lemma}

\begin{proof}
Property \cref{lemma:TNone} follows directly from the definition of $\partsylv{N}$, since  
\[
\partsylv{N}-\upsum{N-1}=\frac{\sylv{N}}{(\sylv{N}-1)^2}\implies 1-\upsum{N-1}/\partsylv{N}=\frac{\sylv{N}}{(\sylv{N}-1)^2}\cdot \partsylv{N}^{-1}\,.\]
 We will prove \cref{lemma:TNtwo} by induction on $k$, starting with $k=N$.
We first convince ourselves that for $k=N$,
\[\frac{\sylv{N}\cdot (\sylv{N}-1)}{\sylv{N}+1}\cdot (1- \upsum{N-1}/\partsylv{N})>\frac{(\sylv{N}-1)^2}{\sylv{N}}\cdot (1-\upsum{N-1}/\partsylv{N})=\partsylv{N}^{-1}
\]
by \cref{lemma:TNone}. Now let $1\leq k<N$ and assume by induction that \cref{lemma:TNtwo} holds for $k+1$. Then
{\allowdisplaybreaks \begin{align*}  
&\frac{\sylv{k}\cdot (\sylv{k}-1)}{\sylv{k}+1}\cdot \bigl(1- \upsum{k-1}/\partsylv{N}\bigr)-\partsylv{N}^{-1}\\
=&\,\frac{\sylv{k}\cdot (\sylv{k}-1)}{\sylv{k}+1}\cdot \left(1- \upsum{k-1}/\partsylv{N}-\frac{\sylv{k}+1}{\sylv{k}\cdot (\sylv{k}-1)}\cdot \partsylv{N}^{-1}\right)\\
=&\,\frac{\sylv{k}\cdot (\sylv{k}-1)}{\sylv{k}+1}\cdot \left(1- \upsum{k-1}/\partsylv{N}-\frac{1}{\sylv{k}-1}\cdot \partsylv{N}^{-1}-\frac{1}{\sylv{k}\cdot (\sylv{k}-1)}\cdot \partsylv{N}^{-1}\right)\\
=&\,\frac{\sylv{k}\cdot (\sylv{k}-1)}{\sylv{k}+1}\cdot \left(1- \left(\upsum{k-1}+\frac{1}{\sylv{k}-1}\right)\cdot\partsylv{N}^{-1}-\frac{1}{\sylv{k}\cdot (\sylv{k}-1)}\cdot \partsylv{N}^{-1}\right)\\
=&\,\frac{\sylv{k}\cdot (\sylv{k}-1)}{\sylv{k}+1}\cdot \left(1- \upsum{k}/\partsylv{N}-\frac{1}{\sylv{k}\cdot(\sylv{k}-1)}\cdot \partsylv{N}^{-1}\right)\\
=&\,\frac{\sylv{k}\cdot (\sylv{k}-1)}{\sylv{k}+1}\cdot \bigl(1- \upsum{k}/\partsylv{N}\bigr)-\frac{1}{\sylv{k}+1}\cdot \partsylv{N}^{-1} \\
>&\,\frac{\sylv{k}\cdot (\sylv{k}-1)}{\sylv{k}+1}\cdot \frac{\sylv{k+1}+1}{\sylv{k+1}\cdot (\sylv{k+1}-1)}\cdot \partsylv{N}^{-1}-\frac{1}{\sylv{k}+1}\cdot \partsylv{N}^{-1}\quad \text{(by induction)}\\
=&\, \partsylv{N}^{-1}\cdot\left(\frac{\sylv{k+1}-1}{\sylv{k}+1}\cdot \frac{\sylv{k+1}+1}{\sylv{k+1}\cdot (\sylv{k+1}-1)}-\frac{1}{\sylv{k}+1}\right)
\quad \text{(by \Cref{lemma:sylvester}\cref{lemmitem:sylvproduct})}\\ 
=&\,\partsylv{N}^{-1}\cdot \left(\frac{\sylv{k+1}+1}{(\sylv{k}+1)\cdot \sylv{k+1}}-\frac{1}{\sylv{k}+1}\right)\\
=&\,\frac{\partsylv{N}^{-1}}{(\sylv{k}+1)\cdot\sylv{k+1}}> 0\,, 
\end{align*}}
which means that
\[\frac{\sylv{k}\cdot (\sylv{k}-1)}{\sylv{k}+1}\cdot \bigl(1- \upsum{k-1}/\partsylv{N}\bigr)>\partsylv{N}^{-1}\,.\]

By induction, property \cref{lemma:TNtwo} holds for all $1\leq k\leq N$.
\end{proof}

Before we prove the competitivity of \genalg, we give one final preparatory lemma.

\begin{lemma}
\label{lemma:densestitem}
Let $A$ be any multiset of items in an instance with $\sum_{x\in A} w(x)\leq 1$. Let $v(A)=\sum_{x\in A} v(x)$ and let $x_d=(w_d,v_d)$ be an item of maximal density in $A$. Then \genalg will achieve a gain of at least
$v(A)\cdot \frac{\lfloor 1/w_d\rfloor}{1/w_d}$ on that instance.
\end{lemma}

\begin{proof}
We know that 
\[v(A)=\sum_{x\in A} v(x)=\sum_{x \in A}\frac{v(x)}{w(x)}\cdot w(x)\leq \sum_{x\in A} \frac{v_d}{w_d}\cdot w(x)\leq\frac{v_d}{w_d}\,.\]
The cumulative value of $x_d$ satisfies
\[v^*(x_d)=v_d\cdot \lfloor 1/w_d\rfloor \geq v(A)\cdot \frac{\lfloor 1/w_d\rfloor}{1/w_d}\;.\]
Since \genalg keeps an item of maximal cumulative value, it will definitely achieve such a gain.
\end{proof}

\begin{theorem}\label{thm:genupper}
\genalg achieves a competitive ratio of at most $\sylvsum< 1.69104$.
\end{theorem}

\begin{proof}
Let $I$ be any non-empty instance and let $\opt\coloneqq\opt(I)$ be an optimal solution on $I$. 
We first start with an intuitive idea of the proof. Ideally, an optimal solution would fill all of its space in the knapsack optimally with the densest possible items. \cref{lemma:densestitem} means that the higher the cumulative weight of the densest item in $\opt$, the better the competitive ratio of \genalg. The smallest cumulative weight that any item can have is at least $1/2$, so we assume this as a worst-case scenario. 
By definition, the gain of \genalg is at least the value of that item. 
Then we can consider the second-densest item in $\opt$. Since it must have weight smaller than $1/2$, it can be packed at least twice and thus its value can be at most $1/2$ the gain of \genalg. 
The smallest cumulative weight such an item can have is at least $2/3$, corresponding to a weight of at least $1/3$, which we again assume as a worst-case scenario.  We then continue to the third-densest item, which must have a weight of at most $1-1/2-1/3=1/6$, and so its value can be at most $1/6$ the gain of \genalg. We can continue this process arbitrarily and see that the value of \opt is at most $1+1/2+1/6+\dots=\sylvsum$ times the gain of \genalg.

More formally, we normalize the values of all items such that $v(\opt)=1$. Let $N\geq 1$. We show that the algorithm has a gain of at least $\partsylv{N}^{-1}$ and thus a competitive ratio of at most $\partsylv{N}$ on $I$. The statement of the theorem then follows since $\partsylv{N}\rightarrow \sylvsum$ as $N\rightarrow \infty$. We first assume that this is not the case, \ie, that the algorithm has a gain of strictly less than $\partsylv{N}^{-1}$. From this, we inductively construct a series of items $y_1=(w_1,v_1),\dots,y_N=(w_N,v_N)$ with the following properties:

\begin{itemize}
\item The optimal solution $\opt$ contains exactly one copy of each item $y_k$.
\item The weight of $y_k$ is bounded by $1/\sylv{k}<w_k\leq 1/(\sylv{k}-1)$ for each  $1\leq k\leq N$.
\item The value of $y_k$ is bounded by $v_k<(\partsylv{N}\cdot(\sylv{k}-1))^{-1}$ for each  $1\leq k\leq N$.
\end{itemize}

We first define $y_1=(w_1,v_1)$ as an item of maximal density in $\opt$. If $w_1\leq 1/2$, then by \Cref{lemma:densestitem,lemma:multiplicity},
\genalg achieves a gain of at least $2/3$. Recall that $\sylv{1}=2$ and $\upsum{0}=0$, so the gain of \genalg is at least
\[2/3=\frac{\sylv{1}\cdot (\sylv{1}-1)}{\sylv{1}+1}\cdot (1- \upsum{0}/\partsylv{N})>\partsylv{N}^{-1}\] by \Cref{lemma:properties}, which we assumed was not the case. This means that $w_1>1/2=1/\sylv{1}$, which also means that the optimal solution can contain only one copy of $y_1$. If $v_1\geq \partsylv{N}^{-1}$, the algorithm achieves a gain of at least $\partsylv{N}^{-1}$, so $v_1<\partsylv{N}^{-1}=(\partsylv{N}\cdot(\sylv{1}-1))^{-1}$, which means that $y_1$ satisfies the properties listed above.

Now assume by induction that we have defined $y_1,\dots,y_{k-1}$ satisfying these conditions for $k\leq N$. 
Since these items are all part of the optimal solution, the other items in $\opt$ must have a combined value of at least $1-\sum_{j=1}^{k-1} v_j>1-\sum_{j=1}^{k-1} (\partsylv{N}\cdot(\sylv{j}-1))^{-1}=1-\upsum{k-1}/\partsylv{N}$. 
They must also have a combined weight of at most $1-\sum_{j=1}^{k-1} w_j<1-\sum_{j=1}^{k-1} 1/\sylv{j}=1/(\sylv{k}-1)$ by \cref{lemma:sylvester}\cref{lemmitem:sylvremainder}. 
This means that it would be possible to pack the combination of all items in $\opt\backslash \{y_1,\dots,y_{k-1}\}$ a total of $\sylv{k}-1$ times for a gain of at least $(\sylv{k}-1)\cdot (1- \upsum{k-1}/\partsylv{N})$. 
Now let $y_k=(w_k,v_k)$ be an item of maximal density among these items, so clearly also $w_k<1/(a_k-1)$. 
By \Cref{lemma:densestitem}, \genalg achieves a gain of at least $\frac{\lfloor 1/w_k\rfloor}{1/w_k} \cdot (\sylv{k}-1)\cdot (1-\upsum{k-1}/\partsylv{N})$. If $w_k\leq 1/\sylv{k}$, then by \Cref{lemma:multiplicity}, its gain is at least $\frac{\sylv{k}}{\sylv{k}+1}\cdot (\sylv{k}-1)\cdot (1- \upsum{k-1}/\partsylv{N})>\partsylv{N}^{-1}$ by \cref{lemma:properties}, which we assumed was not the case. 
Therefore, $w_k>1/a_k$, which also means that the optimal solution can contain only one copy of $y_k$, since it would otherwise contain items of combined weight at least
\[
\left(\sum_{j=1}^{k-1} w_j\right)+2\cdot w_k>\left(\sum_{j=1}^{k-1} \frac{1}{\sylv{j}}\right)+2\cdot\frac{1}{\sylv{k}}
\geq \left(\sum_{j=1}^{k-1} \frac{1}{\sylv{j}}\right)+\frac{1}{\sylv{k}-1}
=1\,,
\]
where the last equality follows from \Cref{lemma:sylvester}\cref{lemmitem:sylvremainder}.

Since $w_k\leq 1/(\sylv{k}-1)$, the item $y_k$ can be packed at least $\sylv{k}-1$ times by itself. Therefore, if $v_k\geq (\partsylv{N}\cdot (\sylv{k}-1))^{-1}$, then \genalg achieves a gain of at least $\partsylv{N}^{-1}$, which we assumed was not the case, so $v_k<(\partsylv{N}\cdot (\sylv{k}-1))^{-1}$. Therefore, $y_k$ satisfies all the properties above. 

This means that we can construct $y_1,\dots,y_N$ satisfying these properties. Further note that in the case $k=N$ we showed that \genalg achieves a gain of at least $\frac{\lfloor 1/w_N\rfloor} {1/w_N}\cdot (\sylv{N}-1)\cdot (1- \upsum{N-1}/\partsylv{N})$. By \Cref{lemma:multiplicity}, this is at least 
\[\frac{\sylv{N}-1}{\sylv{N}}\cdot (\sylv{N}-1)\cdot (1- \upsum{N-1}/\partsylv{N})\,,\]
which is equal to $\partsylv{N}^{-1}$
by \Cref{lemma:properties},
once again contradicting our assumption. We are forced to conclude that this assumption was wrong.
The algorithm \genalg will therefore always achieve a gain of at least $\partsylv{N}^{-1}$ and thus a competitive ratio of at most $\partsylv{N}$.
\end{proof}

We now show that this simple algorithm is in fact the best possible, \ie, that its competitive ratio cannot be improved upon:

\begin{theorem}\label{thm:genlower}
No deterministic algorithm for the \textsc{General Unbounded Knapsack Problem with Removal} can have a better competitive ratio than $\sylvsum$.
\end{theorem}
\begin{proof}
We present an adversarial sequence of items to show that this is impossible. The sequence starts with an item $(1,\upsum{N})$, and then consists of a large number of subsequences. Each subsequence has the same length $N$ and consists of items with weight slightly larger than $1/\sylv{i}$ and value proportional to $1/(\sylv{i}-1)$ for $1\leq i\leq N$. This means that the items in each subsequence have a combined value proportional to $\upsum{N}$. However, in each successive subsequence, the value of each item is multiplied by a very small constant, which slowly increases the value of the optimal solution. If the original item is ever removed, the instance stops, and the single item that can be packed in this step is by itself off by a factor of approximately $\upsum{N}$ from the optimal solution.

More specifically, let \alg be any deterministic algorithm. We show that it has a competitive ratio of at least $\sylvsum-\eps$ for any $\eps>0$. Let $N\in \N$ be large enough that $\upsum{N}>\sylvsum-\eps/2$. Then let $\delta>0$ be small enough that $\upsum{N}/(1+\delta)>\upsum{N}-\eps/2$. Finally, let $M\in \N$ be large enough that $(1+\delta)^M>\sylvsum-\eps$. 

The sequence starts with an item $(1,\upsum{N})$, which we can assume \alg packs, since its competitive ratio is unbounded otherwise. After that, we define the following items: 
\[
x_{i,j}=(w_i,v_{i,j})=\Big(\frac{1}{\sylv{i}}+\eps', \frac{(1+\delta)^j}{\sylv{i}-1}\Big)\quad\text{for }1\leq i\leq N,\, 1\leq j\leq M\,.
\]
Here, $\eps'>0$ is chosen small enough that $\sum_{i=1}^N (1/\sylv{i}+\eps')\leq 1$, which is possible by \cref{lemma:sylvester}\cref{lemmitem:sylvremainder}. Note that this means that for any fixed $j$, the items $x_{i,j}$ can be packed together, for a combined value of $\sum_{i=1}^N v_{i,j}=(1+\delta)^j\cdot \upsum{N}$. The algorithm now receives the items $x_{i,j}$ for $1\leq i\leq N$, for each $1\leq j\leq M$ in order. 

If \alg never removes the first item $(1,\upsum{N})$, then it can pack no other items. The optimal solution has value at least $\sum_{i=1}^N v_{i,M}=(1+\delta)^M\cdot \upsum{N}$.
The competitive ratio of \alg is therefore at least $(1+\delta)^M>\sylvsum-\eps$.

So assume \alg removes the item $(1,\upsum{N})$ to pack an item $x_{i,j}$. In this case, the instance immediately ends, and \alg's gain is at most $(1+\delta)^j$. The optimal solution has value at least $(1+\delta)^{j-1}\cdot \upsum{N}$: If $j=1$, its value is at least the value $\upsum{N}$ of the first item. If $j>1$, its value is at least $\sum_{i=1}^N v_{i,j-1}=(1+\delta)^{j-1}\cdot \upsum{N}$. This means that the competitive ratio of \alg is at least
$\upsum{N}/(1+\delta)>\upsum{N}-\eps/2>\sylvsum-\eps$. 

In either case, we showed that \alg's competitive ratio is at least $\sylvsum-\eps$. Since this holds for any $\eps>0$, it cannot have a competitive ratio of less than $\sylvsum$.
\end{proof}


\begin{thebibliography}{10}

\bibitem{BKKR2014}
Hans-Joachim B{\"{o}}ckenhauer, Dennis Komm, Richard Kr{\'{a}}lovi{\v{c}}, and
  Peter Rossmanith.
\newblock The online knapsack problem: Advice and randomization.
\newblock {\em Theoretical Computer Science}, 527:61--72, 2014.
\newblock \href {https://doi.org/10.1016/j.tcs.2014.01.027}
  {\path{doi:10.1016/j.tcs.2014.01.027}}.

\bibitem{BGHKKLMRS2024}
Hans-Joachim Böckenhauer, Matthias Gehnen, Juraj Hromkovi\v{c}, Ralf Klasing,
  Dennis Komm, Henri Lotze, Daniel Mock, Peter Rossmanith, and Moritz Stocker.
\newblock Online unbounded knapsack.
\newblock {\em Theory of Computing Systems}, 69(1):1--25, 2025.
\newblock \href {https://doi.org/10.1007/s00224-025-10215-0}
  {\path{doi:10.1007/s00224-025-10215-0}}.

\bibitem{BHKRS2026}
Hans-Joachim Böckenhauer, Juraj Hromkovič, Dennis Komm, Peter Rossmanith, and
  Moritz Stocker.
\newblock A survey of online knapsack problems.
\newblock {\em Discrete Applied Mathematics}, 378:492--507, 2026.
\newblock \href {https://doi.org/10.1016/j.dam.2025.08.011}
  {\path{doi:10.1016/j.dam.2025.08.011}}.

\bibitem{CJS2016}
Marek Cygan, {\L}ukasz Je\.{z}, and Ji\v{r}{\'{\i}} Sgall.
\newblock Online knapsack revisited.
\newblock {\em Theory of Computing Systems}, 58(1):153--190, 2016.
\newblock \href {https://doi.org/10.1007/s00224-014-9566-4}
  {\path{doi:10.1007/s00224-014-9566-4}}.

\bibitem{GS2026}
Matthias Gehnen and Moritz Stocker.
\newblock Stealing from the dragon's hoard: Online unbounded knapsack with
  removal.
\newblock In Meena Mahajan, Florin Manea, Annabelle McIver, and Kim~Thang
  Nguyen, editors, {\em 43rd International Symposium on Theoretical Aspects of
  Computer Science ({STACS} 2026)}, volume 364 of {\em LIPIcs}, pages
  43:1--43:17. Schloss Dagstuhl - Leibniz-Zentrum f{\"{u}}r Informatik, 2026.
\newblock \href {https://doi.org/10.4230/LIPICS.STACS.2026.43}
  {\path{doi:10.4230/LIPICS.STACS.2026.43}}.

\bibitem{Haechler2025}
Valentin H\"{a}chler.
\newblock Untrusted predictions for knapsack problem variants.
\newblock Bachelor's Thesis, ETH Zurich, 2025.
\newblock \href {https://doi.org/10.3929/ethz-c-000782659}
  {\path{doi:10.3929/ethz-c-000782659}}.

\bibitem{HKM2015}
Xin Han, Yasushi Kawase, and Kazuhisa Makino.
\newblock Randomized algorithms for online knapsack problems.
\newblock {\em Theoretical Computer Science}, 562:395--405, 2015.
\newblock \href {https://doi.org/10.1016/j.tcs.2014.10.017}
  {\path{doi:10.1016/j.tcs.2014.10.017}}.

\bibitem{HKMY2019}
Xin Han, Yasushi Kawase, Kazuhisa Makino, and Haruki Yokomaku.
\newblock Online knapsack problems with a resource buffer.
\newblock In Pinyan Lu and Guochuan Zhang, editors, {\em 30th International
  Symposium on Algorithms and Computation ({ISAAC} 2019)}, volume 149 of {\em
  LIPIcs}, pages 28:1--28:14. Schloss Dagstuhl - Leibniz-Zentrum f{\"{u}}r
  Informatik, 2019.
\newblock \href {https://doi.org/10.4230/LIPIcs.ISAAC.2019.28}
  {\path{doi:10.4230/LIPIcs.ISAAC.2019.28}}.

\bibitem{IT2002}
Kazuo Iwama and Shiro Taketomi.
\newblock Removable online knapsack problems.
\newblock In Peter Widmayer, Francisco~Triguero Ruiz, Rafael~Morales Bueno,
  Matthew Hennessy, Stephan~J. Eidenbenz, and Ricardo Conejo, editors, {\em
  29th International Colloquium on Automata, Languages and Programming ({ICALP}
  2002)}, volume 2380 of {\em Lecture Notes in Computer Science}, pages
  293--305. Springer, 2002.
\newblock \href {https://doi.org/10.1007/3-540-45465-9\_26}
  {\path{doi:10.1007/3-540-45465-9\_26}}.

\bibitem{IZ2010}
Kazuo Iwama and Guochuan Zhang.
\newblock Online knapsack with resource augmentation.
\newblock {\em Information Processing Letters}, 110(22):1016--1020, 2010.
\newblock \href {https://doi.org/10.1016/j.ipl.2010.08.013}
  {\path{doi:10.1016/j.ipl.2010.08.013}}.

\bibitem{Komm2016}
Dennis Komm.
\newblock {\em An Introduction to Online Computation -- Determinism,
  Randomization, Advice}.
\newblock Springer, 2016.
\newblock \href {https://doi.org/10.1007/978-3-319-42749-2}
  {\path{doi:10.1007/978-3-319-42749-2}}.

\bibitem{LL1985}
C.~C. Lee and D.~T. Lee.
\newblock A simple on-line bin-packing algorithm.
\newblock {\em J. {ACM}}, 32(3):562--572, 1985.
\newblock \href {https://doi.org/10.1145/3828.3833}
  {\path{doi:10.1145/3828.3833}}.

\bibitem{MV1995}
Alberto Marchetti{-}Spaccamela and Carlo Vercellis.
\newblock Stochastic on-line knapsack problems.
\newblock {\em Mathematical Programming}, 68:73--104, 1995.
\newblock \href {https://doi.org/10.1007/BF01585758}
  {\path{doi:10.1007/BF01585758}}.

\bibitem{oeis}
{OEIS Foundation Inc.}
\newblock Entry {A}000058 in {T}he {O}n-{L}ine {E}ncyclopedia of {I}nteger
  {S}equences, 2026.
\newblock URL: \url{https://oeis.org/A000058}.

\bibitem{Sylvester1880}
J.~J. Sylvester.
\newblock On a point in the theory of vulgar fractions.
\newblock {\em American Journal of Mathematics}, 3(4):332--335, 1880.
\newblock \href {https://doi.org/10.2307/2369261} {\path{doi:10.2307/2369261}}.

\bibitem{Yao1977}
Andrew C.-C. Yao.
\newblock Probabilistic computations: Toward a unified measure of complexity
  (extended abstract).
\newblock In {\em Proceedings of the 18th Annual Symposium on Foundations of
  Computer Science ({FOCS} 1977)}, pages 222--227, 1977.
\newblock \href {https://doi.org/10.1109/SFCS.1977.24}
  {\path{doi:10.1109/SFCS.1977.24}}.

\end{thebibliography}
\end{document}